\documentclass{aa}
\usepackage{epsfig}
\begin{document}
\title{Multiple Timescales in Cataclysmic Binaries}
 \subtitle{The Low-Field Magnetic Dwarf Nova DO Dra
\fnmsep\thanks{Table 3 is only available in electronic form at http://www.edpsciences.org}
}
   \author{I.L. Andronov\inst{1,2} \and L.L.Chinarova\inst{2} \and W. Han\inst{3} \and Y. Kim\inst{4,}\inst{5} \and J.-N. Yoon\inst{4}
         }
   \institute{
Odessa National Maritime University,
             Mechnikov str., 34, 65029, Odessa, Ukraine,\\
 \email{tt\_ari@ukr.net, andronov@osmu.odessa.ua, il-a@online.ua, uavso@pochta.ru, bfcyg@mail.ru}
         \and
Astronomical Observatory, Odessa National University,
             T.G.Shevchenko Park, 65014, Odessa, Ukraine,
         \and
Korea Astronomy Observatory and Space Science Institute, Daejeon 305-348, Korea
         \and
University Observatory, Chungbuk National University, 361-763,
Cheongju, Korea
         \and
Institute for Basic Science Research, Chungbuk National
University, 361-63, Korea
             }
\date{Received November 13, 2007 / Accepted January 17, 2008}
\abstract{}{We study the variability of the cataclysmic variable DO Dra, on time-scales of between minutes and decades.}
 {The observations were obtained at the Korean 1m telescope at the Mt. Lemmon in 2006-2007, 14 observational runs cover 45 hours. The table of individual observations is available electronically. Additionally, we have used 1509 patrol observations from the international AFOEV and VSOLJ databases.} {The characteristic decay time dt/dm=0.902(3) days/mag was estimated from our 3 nights of CCD R observations, which cover the descending branch of the outburst in 2006. The range of the outburst cycle  is from $311^{\rm d}$ to $422^{\rm d},$ contrary to a previous estimate of $870^{\rm d}.$ The ``quiescent" data show a photometric wave with a cycle $\sim 303(15)^{\rm d}.$
We analyzed the profile of the "composite" (or "mean") outburst. We discovered however, that a variety of different outburst heights and durations had occurred, contrary to theoretical predictions. The analysis of the historical data has shown a correlation between the decay time dt/dm and the outburst maximum brightness with a slope d(dt/dm)/dm=0.37(9).
With increasing maximum brightness, we find that the decay time also increases; this is in contrast to the model predictions, which indicate that outbursts should have a constant shape. This is interpreted as representing the presence of outburst-to-outburst variability of the magnetospheric radius. A presence of a number of missed weak narrow outbursts is predicted from this statistical relationship.
We tabulate characteristics of the "quasi-orbital" variations, which indicate that an amplitude maximum occurs between quiescence and the outburst peak.
The semi-amplitude of the spin variability does not exceeded 0.02 mag. A new type of variability is detected, during 3 subsequent nights in 2007: periodic (during one nightly run) oscillations with rapidly-decreasing frequency from 86 to 47 cycles/day and a semi-amplitude increasing from $0\fm06$ to $0\fm10,$ during a monotonic brightness increase from $14\fm27$ to $14\fm13.$ This phenomenon was observed only during an unusually prolonged event of $\sim1$ mag brightening in 2007 (lasting till autumn), during which no (expected) outburst was detected. We refer to this behaviour as to the ``transient periodic oscillations" (TPO).

We attribute the frequency decrease to "beat"-type of the variability, probably caused by irradiation of a cloud that is spiralling down to the white dwarf. Its frequency would then increase and coverge towards the spin frequency. To study this new and interesting phenomenon, new regular photometric and spectral (in a "target of opportunity" mode) observations are required.
}{}
\authorrunning{Andronov et al.}
\titlerunning{Magnetic Dwarf Nova DO Dra}
\maketitle


\section{Introduction}
DO Dra belongs to a class of cataclysmic variables that have accretion quite unaffected by the magnetic field. Such objects are called ``intermediate polars" or ``DQ Her - type stars" (see Patterson (1994), Warner (1995), Norton et al. (2004) and Hellier (2001) for more detailed description).

This object was detected as an X-ray source 2A 1150+720 and later classified as a cataclysmic variable star by Patterson et al. (1982). It was also detected as a cataclysmic variable in the Palomar-Green Survey, and listed as PG 1140+719 (Green et al. 1982).
They suggested an identification of this object with a previously-registered variable YY Dra.

The designation "YY Dra" was however assigned to an eclipsing variable that had a brightness range of 12\fm9 -- $<14\fm5$, and period 4\fd21123 as measured by Tsesevich (1934), and was almost coincident in co-ordinates with the X-ray source.
Possibly due to a misprint of coordinates, the ``true" (eclipsing) YY Dra was not found until now. Wenzel (1983) failed to find variability of the $12^{\rm m}$ star close to the published position of YY Dra, but found an eruptive object at the position of  PG 1140+719. He detected this object on only two plates from 700, and thus classified the object as a dwarf nova with an extremely long cycle length. This is not an eclipsing variable with a well-defined period, and thus a separate official GCVS name ``DO Dra" was assigned to PG 1140+719 (Kholopov et al. 1985, Samus' et al., 2007). The designation of the star was discussed by Patterson and Eisenman (1987) and Kholopov and Samus (1987). In the literature, DO Dra can still be referred to as "YY Dra" or "DO/YY Dra".

The physical nature of this object may be inferred from the photometric behaviour of the system as of the dwarf nova. These systems are close binaries with a red dwarf filling its Roche lobe and a white dwarf. The plasma stream from the secondary forms an accretion disk, which becomes cyclically unstable after reaching some critical viscosity (cf. Warner 1995). Smak (1984), based on theoretical models, distinguished between two types of dwarf nova outbursts - with an onset in the {\it outer} (type A) and {\it inner} (type B) parts of the accretion disk.

An unusual feature of DO Dra is a short outburst as compared to other systems. This is interpreted by a relatively large inner radius of the accretion disk, which, contrary to ``non-magnetic" dwarf novae, is equal to the Alfven radius $R_A,$ rather than to the radius of the white dwarf (cf. Angelini and Verbunt 1989). The detailed study of the long-term and outburst behaviour of DO Dra was presented by \v{S}imon (2000) on the base of visual amateur observations. He estimated an outburst cycle length of $\sim868^{\rm d},$ but further monitoring presented in the international databases of AFOEV (2007) and VSOLJ (2006) imply much shorter time intervals (as discussed in Section 2).

So large outburst cycle is in addition implied by the relatively large inner radius $R_{in}$ of the accretion disk. For non-magnetic cataclysmic variables, $R_{in}$ is compared with the radius of the white dwarf $R_{wd}.$
For intermediate polars $R_{in}\underline{\sim} R_{A},$ where $R_{A}$ is the radius of magnetosphere (Angelini \& Verbunt 1989). Another bright magnetic dwarf nova with short outbursts and long outburst cycle is GK Per (see \v{S}imon 2007 for a recent review).
Cannizzo \& Mattei (1992, 1998) have thoroughly studied characteristics of 705 outbursts of the prototype dwarf nova SS Cyg observed during 95 years. They found a bimodal distribution. The slope dm/dt for the descending branch (the decay of the outburst) is nearly constant (within a dozen per cent) for different outbursts, in agreement with theoretical models.

Patterson et al. (1992) found a ``fundamental" period of $550^s\pm3^s$ with a double-peak structure of the phase curve. The highest peak in the amplitude spectrum therefore occurs at a half of the period and has an amplitude equal to $275^s\pm1^s.$ For some nights, the shorter period was found to be $266^s.$
This value was interpreted as a half of the siderial period of the magnetic white dwarf.
This interpretation has been made in other papers, even though it is accepted that the "fundamental" spin period was determined as $P_{spin}=529.31(2)$ s (Haswell et al. 1997). The number in parentheses corresponds to an accuracy estimate (i.e. standard error). Haswell et al. (1997) precisely determined the orbital period $P_{orb}=0\fd16537398(17),$ the initial epoch $T_0=244683.4376(5)$ for the inferior conjunction of the secondary (we use these values here to compute the orbital phases), the masses of both the secondary $(M_2=0.375(14)M_\odot,$ and the white dwarf $M_{wd}=0.83(10)M_\odot,$ and the inclination angle $i=45^\circ(4^\circ).$

With such a moderate inclination, there are no eclipses of the accretion disk/columns by the red dwarf, so the variability with orbital phase may be attributed mainly to the secondary star - either the ellipticity effect (cf. detection for EF Eri by Allen and Cherepashchuk (1982)) or irradiation (cf. Basko and Sunyaev 1973, King and Lasota 1984).
Results of X-ray/optical studies of two outbursts (1999 and 2000) were presented by Szkody et al. (2002), who noted
that the behaviour of the spin pulse amplitude with luminosity was unusual.

Norton et al. (1999) have split the group of intermediate polars into two subclasses with relatively large and relatively weak magnetic field and classified DO Dra as belonging to the second group (with DQ Her, V709 Cas et al.). In the catalogue of Ritter and Kolb (2003), there are 48 objects classified as intermediate polars (IP) or DQ Her-type stars (DQ), from which 36 (75\%) are nova-like (NL) variables, in 7 (15\%), the Nova outbursts were detected, and 5 (10\%) show dwarf nova-type outbursts. Consequently, the latter group may be called either "magnetic dwarf novae", or "outbursting intermediate polars".

In this paper, we study the time and luminosity variability in observations of DO Dra, acquired over 13 nights, at quiescence, outburst peak, and at the descending branch of the light curve.
Additionally, we reanalyze patrol observations published in the AFOEV (2007) and VSOLJ (2006) databases.

\begin{figure}
\psfig{file=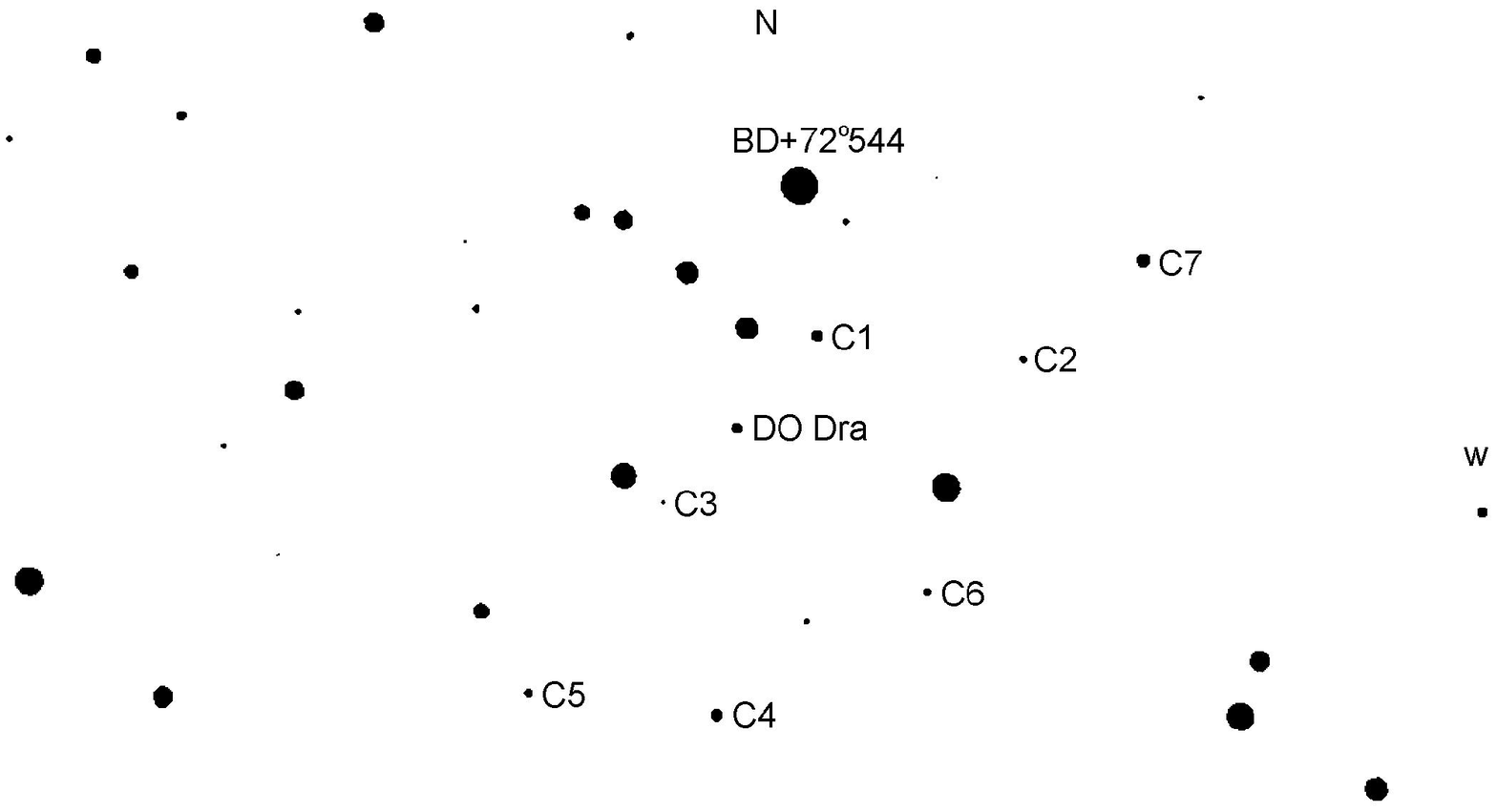} \caption{Finding chart for DO Dra. The
size of the field is $7.3'\times4'.$}
\end{figure}


\begin{table}
\label{t1}
\caption{Mean brightness of DO Dra and the comparison stars from the Mt.Lemmon observations (R).
The values of $\sigma$ correspond to the r.m.s. scatter of a single observation in respect to the "artificial comparison star". The statistical accuracy of the mean value is by a factor of $n^{1/2}=1350^{1/2}\approx37$ times smaller, i.e. does not exceed 0\fm001.
}
\centerline{\begin{tabular}{cccccccccc}
\hline
Star&R&$\sigma_R$&Star&R&$\sigma_R$\\
\hline
DO &13.747 & 1.320 & C4 &13.754 & 0.024 \\
C1 &13.840 & 0.014 & C5 &14.631 & 0.033 \\
C2 &15.016 & 0.020 & C6 &14.831 & 0.026 \\
C3 &15.378 & 0.024 & C7 &13.543 & 0.009 \\
\hline
\end{tabular}}
\end{table}

\section{Observations and Comparison Stars}

\subsection{Observations}

R-band time-series observations were acquired using a 2K by 2K CCD camera mounted on the LOAO 1.0m telescope, Arizona. Given the CCD plate scale of 0.64 arcseconds/pixel at the f/7.5 Cassegrain focus, the image field-of-view was 22.2 arcminutes by 22.2 arcminutes.

Our image data reduction was completed using bias, dark and flat-field calibration data, and the IRAF package CCDRED. Instrumental stellar magnitudes were derived empirically by fitting the point-spread functions (PSFs) of stars using the IRAF package DAOPHOT (Stetson 1987; Massey
\& Davis 1992).

The journal of observations is presented in Table 2.
During this campaign, we obtained 1511 observations over 45 hours in 14 nights from October 29, 2005 to March 3, 2007.

\begin{table}
\label{t2}
\caption{Journal of observations of DO Dra:
Time of the begin $t_b$ and end $t_e$ (in HJD-2400000) of observations (for the majority of nights, the observations ended on the next integer JD), the integer part of the starting Julian date JD is used for a legend of the run, e.g. 52752);
number of observations $n$; magnitude range for individual data points $m_{max}$, $m_{min}$;
nightly mean $\langle m\rangle$ and it's accuracy estimate;
r.m.s. deviation of the single observation from the mean $\sigma(m);$
exp - exposure in seconds.
}
\begin{tabular}{crcccr}
\hline
$t_b-t_e$&~~$n$&range& $\langle m\rangle$&$\sigma(m)$&exp\\
\hline
53672.996-.031& 38& 14.20-14.55& 14.363(17)&  0.105&   79\\
53752.898-.071& 91& 14.97-15.22& 15.126(06)&  0.055&  166\\
53753.876-.055& 78& 15.07-15.35& 15.227(09)&  0.076&  166\\
53773.861-.066&228& 11.24-11.74& 11.532(06)&  0.083&   71\\
53774.848-.068&201& 12.34-12.85& 12.599(07)&  0.097&   85\\
53776.957-.067& 90& 14.68-15.12& 14.891(11)&  0.101&  105\\
53777.878-.065&153& 14.82-15.18& 15.026(07)&  0.085&  105\\
53780.841-.848&  6& 15.01-15.24& 15.151(34)&  0.084&  106\\
53782.030-.065& 30& 15.24-15.56& 15.361(11)&  0.058&  105\\
53786.885-.053&136& 15.07-15.36& 15.215(05)&  0.061&  105\\
53787.878-.963& 70& 14.91-15.18& 15.024(08)&  0.070&  105\\
54161.889-.001& 60& 14.10-14.49& 14.269(11)&  0.086&  150\\
54162.879-.036&153& 13.81-14.61& 14.221(12)&  0.150&   83\\
54163.855-.036&177& 13.77-14.47& 14.127(10)&  0.130&   84\\
\hline
\end{tabular}
\end{table}

The original observations (HJD, magnitude) are presented in Table 3 (electronically only).

\begin{table}
\caption{Table of R observations (HJD-2400000, magnitude) obtained at the 1m telescope of the MtLemmon observatory, Korea. {\em The table is very long, contains 1511 lines and is planed to be published as a text file electronically only}}
\end{table}

\subsection{Comparison stars}
Photometric BV standards for our field-of-view were published by Henden and Honeycutt (1995). The chart based on Hipparcos and Tycho catalogues is presented by the VSNET (2005) also in B and V. The finding chart is shown in Fig. 1. There is a nearby bright star BD$+$72$^\circ$544 with B=11\fm02, V=9\fm65. The R magnitudes of some comparison stars have been published by Cook (2005). The apparent magnitudes of the ``main" comparison star C1 are B=15\fm082, V=14\fm28 (Henden and Honeycutt 1995), R=13\fm84 (Cook 2005). We used for the calibration of our observations the latter value R=13\fm84.

We list the magnitudes of other comparison stars in Table 3. These magnitudes were determined using the "artificial" (mean-weighted) star method (Kim et al. 2004), implemented by the program "MCV" (Andronov and Baklanov 2004). The error of the brightness estimate of the artificial comparison star is 0\fm0061, which is much lower than that of any error of magnitudes of the comparison stars. The smallest magnitude errors for the actual comparison stars were C1 (0\fm014) and C7(0\fm009). Obviously, the variable shows much larger value of the root-mean-squared deviation of the mean (1\fm32) than any of the comparison stars, becaues of its intrinsic variability.
The typical RMS errors for a single magnitude measurement of the variable, ranges from 0\fm008 (at the outburst peak), to 0\fm026 (during the 6 nights of the faint state).

\section{Dwarf Nova Outbursts}

\begin{figure*}
\psfig{file=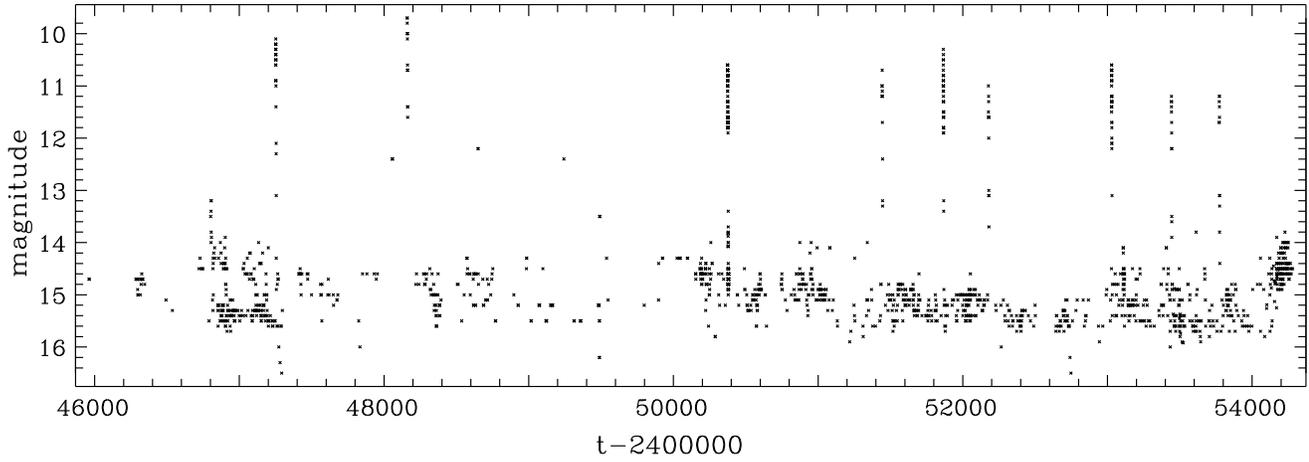}
\label{f2}
\caption{The historical light curve based on observations published in the AFOEV (2007) and VSOLJ (2007) databases. Only 1509 "sure" observations are shown, without "fainter than" or "unsure" ones.}
\end{figure*}

\begin{figure*}
\psfig{file=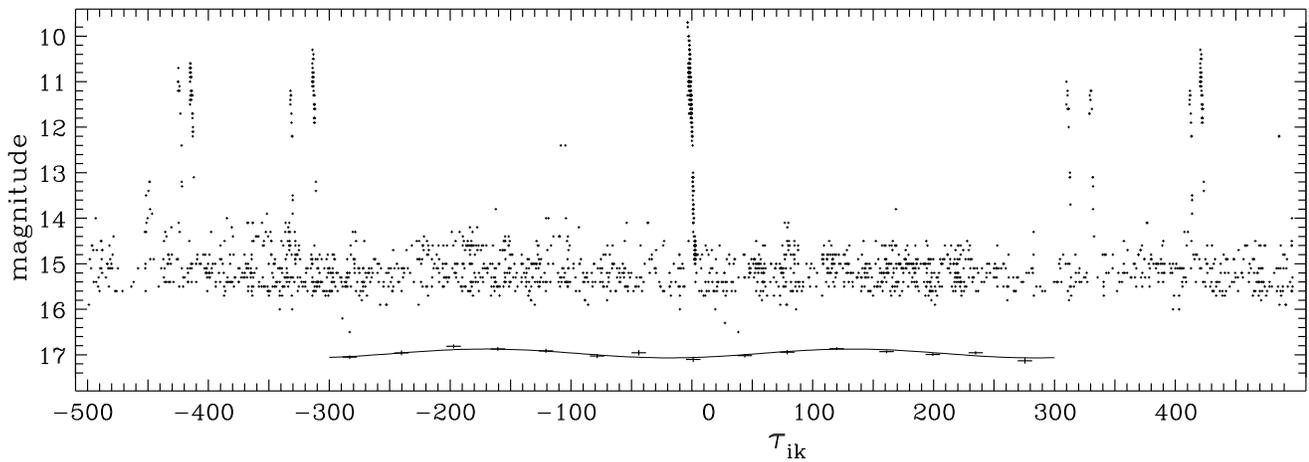}
\label{f3}
\caption{The composite light curve based on the AFOEV and VSOLJ international databases. The abscissa shows differences $\tau_{ik}=t_i-t_{o,k},$ where $t_{o,k}$ is time of crossing the fixed brightness level $V=12^m$ for $k^{th}$ outburst. The bottom line corresponds to a part of a periodic $(P\approx303^{\rm d})$ sine fit to the "out of outburst" observations for $-300^{\rm d}<\tau_{ik}<+300^{\rm d}.$ The $0^{\rm d}$ mean values are shown with the corresponding error bars.}
\end{figure*}

Wenzel (1983) have found 2 outbursts from 700 Sonneberg plates taken in 1963-1982, and suggested a very long outburst cycle of 5-20 years. Hazen (1986) detected 3 outbursts on 441 Harvard plates taken in 1890-1984
and noticed a small duration of the outburst of $<10^{\rm d}$ or ``probably even less" and thus suggested that many outbursts are missing. Lange (2007) lists 16 registered outbursts since 1936 with a total duration of 3-5 days. The last outburst was registered on February 5, 2006 and after 4 days the star was again in a quiet state. In an addition to our own observations (3 nights during this outburst and 10 nights ``out of outburst"), the historical light curve has been analyzed. The visual observations published in the databases of AFOEV (2007) and VSOLJ (2007) have been used. The last update of the databases was on October 27, 2007. From a total number of 13785 observations, we analyzed only 1509, having removed observations flagged as "unsure", "fainter than" and "CCD". The resulting light curve is shown in Fig.2.

\subsection{Outburst Cycle}

To compare different outbursts, we have composed a cumulative light curve by using the relative time $\tau_{ik}=t_i-t_{o,k},$ instead of the original times $t_i,$ where $t_{o,k}$ is the moment of time, when the light curve crosses the fixed brightness level $V=12^m$ during the outburst decay (number $k$). This approach follows that of \v{S}imon (2000). Originally we defined $t_{o,k}$ to be the outburst time. However, because of missing ``true" outburst maxima, the time of crossing is often more accurately determined, e.g. using linear interpolation of neighbouring points.

It is a modification of the phase curve for a case of aperiodic variations. In practice, we have determined the moments $t_{o,k}$ for clearly-defined outbursts, and then computed $\tau_{ik}$ and removed the ``far" data outside the interval $-\delta\tau<\tau_{ik}<\delta\tau.$ The corresponding light curve is shown in Fig. 3 for $\delta\tau=500^d.$ The time interval between the outbursts is not constant, thus the outbursts occur at different relative times. Some outbursts are present three times at the same graph, because the preceeding outburst is shown at negative $\tau_{ik},$ and then (for next value of $k$) the same outburst becomes a central one. But it may be a succeeding one, so may occur asymmetrically at positive $\tau_{ik}.$ Generally, the preceeding and succeeding time intervals are different, so the intervals between the curves, which correspond to a same outburst, are also different. However, at the place corresponding to some $(-\tau_{ik}),$ may occur another outburst from a pair.

In a case of truly periodic variations of period $P,$ there should be repetitive outbursts occurring at $\tau_{ik}=mP,$ where $m$ is an arbitrary integer.

The group of outbursts at Fig.3 at the abscissa $\tau_{ik},$ which is closest to the central one, corresponds to  $|\tau_{ik}|$ from 311 to 422 days, which may be an estimate for an outburst cycle.
The second group of detected outbursts are found at $|\tau_{ik}|$ in the range of 734-908 days (not shown in Fig. 3). \v{S}imon (2000) argued for a cycle of $868^{\rm d},$ which lies in this range. However, further more dense observations showed at least 4 smaller intervals between outbursts, as seen from Fig. 2 and 3. Moreover, apparently, there are pairs of "$400^{\rm d}; 300^{\rm d}$" outburst cycles in Fig.3. So the published larger estimate of  $868^{\rm d}$ possibly corresponds to "multiple" cycle length.

Two outlying observational data-points at $\tau_{ik}=-105^{\rm d}$ and $-108^{\rm d})$, acquired by observer P. Schmeer, remain unconfirmed by independent observations. Furthermore, the measured magnitudes for these data are systematically fainter than is typical for DO Dra during outbursts.

We agree with independent sets of observations, completed by this same observer, and therefore decide to consider these data further here. If we assume that these data correspond to a real outburst, then the typical cycle period must be reduced to approximately $100^d.$ Furthermore, this would imply that eighty percent of outbursts have been undetected by completed observations. This seems incredible, even though the typical duration of outbursts is a few days, and it would be difficult to arrange all-year monitoring.

Another explanation of these points may be short faint outbursts. For comparison, a bimodal distribution of the outburst durations was found by Cannizzo \& Mattei (1992) in SS Cyg. In this case, in an addition to previously detected ``long" outbursts, there may take place ``short" (a couple of days) outbursts. We propose to check this idea during further monitoring observations.

The ``out of outburst" (or ``quiescent luminosity state") observations show fluctuations on the timescale of tens to hundreds of days (cf. \v{S}imon 2000, 2007). To study this phenomenon as a function of time difference from the outburst $\tau_{ik},$ we have analyzed the composite light curve, as described above.
The periodogram analysis of the quiescent data (i.e. in the range $(-300^{\rm d},-6^{\rm d}),(6^{\rm d},300^{\rm d}))$ has been completed using the program "Fo" (Andronov 1994). It shows the highest peak at the period $P=303\pm15$ with a corresponding semi-amplitude $97\pm16$ mmag (i.e. $6\sigma$), mean magnitude $15\fm17\pm0\fm01$ and moment of minimum at $\tau_{ik}=19^{\rm d}\pm7^{\rm d}$. The false alarm probability (FAP) is $10^{-5.5},$ so the wave appears to be statistically significant. However, we do not assume this type of variability to be strictly periodic. The coincidence of the "period" (or, expectedly, "cycle") with the semi-width of the interval (which is chosen to be slightly smaller than the expected outburst cycle) may argue for variations of the mean brightness between outbursts.

The second peak in the periodogram occurs at $39\fd9\pm0\fd3$ with a smaller amplitude of $67\pm15$ mmag, at the limit of detection (FAP=$10^{-2}$). It is distinctly longer than one month, so this should not be a selection effect. In Fig.3,
one may see variations at this timescale. However, there is no known type of periodic variability, so we do not expect a ``true" periodicity, and just mention some variability at this timescale.

 The mean values computed for $40^{\rm d}$ intervals of $\tau_{ik}$ range from $15\fm01\pm0\fm01$ to $15\fm30\pm0\fm06,$ i.e. the "peak-to-peak" amplitude exceeds $4.1\sigma.$

The overall amplitude of mean brightness variations during quiescence reaches $0\fm4.$ This value is smaller than that (1\fm0, range 14\fm8-15\fm8) of \v{S}imon (2000, 2007), arguing that aperiodic fluctuations are of larger amplitude than the cyclic ones.

An important correlation between the outburst cycle and the maximum brightness was found by Cannizzo \& Mattei (1992, 1998), and also discussed by Cannizzo (1993).
There should be an inverse correlation between the quiescent brightness and the recurrence time of outbursts. This is because the instability is triggered after a critical amount of material has accumulated, a more rapid accumulation implies a shorter time between outbursts, and part of the luminosity at quiescence is generated by the hot spot reflecting the impact of the mass stream on the outer disk edge.

In addition, the decay time of the model is affected by the cooling-front transition, which begins at the outer disk edge and moves inwards. The rise time, in contrast, can be either rapid or slow depending on whether the instability begins in the outer or inner disk.

Thus there is a model prediction of a stable dt/dm value from outburst-to-outburst, within a given system. Such a result was found for a much brighter dwarf nova SS Cygni.

We are unable, however, to achieve a similar result for DO Dra, because of the lack of detected outbursts.

For DO Dra, a number of detected outburst maxima is small, so probably there are many missing outbursts, and, at the moment, it is not possible to get statistically significant results. Continuation of monitoring, also to search for such correlation, is an important task for further study of the magnetic white dwarf nova DO Dra.

Another interesting phenomenon is an unusually high peak of brightness up to $\sim14\fm5,$ observed in the right half of Fig. 2 (summer of 2007), and confirmed by the observations of BAVR (Lange 2007). Its full width at half-maximum is $\sim 150^{\rm d},$ an excellent agreement with the $300^{\rm d}$ cycle. However, no narrow outburst was detected during this brightening, as would be expected for a suggested cycle length $300^{\rm d}-400^{\rm d}.$

Variations of the accretion rate caused by the solar-type variability of the red dwarf are characteristized by a timescale of few years (cf. Bianchini 1990, Richman et al. 1994).

Luminosity in the quiescent state varies on the timescale of one year, which is short to be explained by the solar-type activity. Because this timescale is close to the cycle period, must be related to the outburst mechanism. We argue that regular monitoring even between outbursts, is required.

\subsection{Mean Outburst Profile}

\begin{figure}
\psfig{file=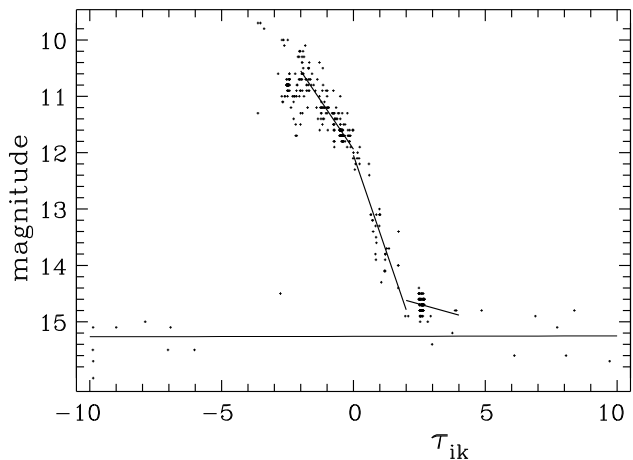}
\label{f4}
\caption{The composite light curve based on the AFOEV and VSOLJ international databases near the outburst. The inclined lines at the outburst show linear fits to the observations in the intervals $(-2^{\rm d},0^{\rm d}),$ $(0^{\rm d},2^{\rm d})$ and $(2^{\rm d},4^{\rm d}),$ respectively. The bottom line corresponds to a part of a periodic $(P\approx303^{\rm d})$ sine fit to the "out of outburst" observations.}
\end{figure}

\begin{figure}
\psfig{file=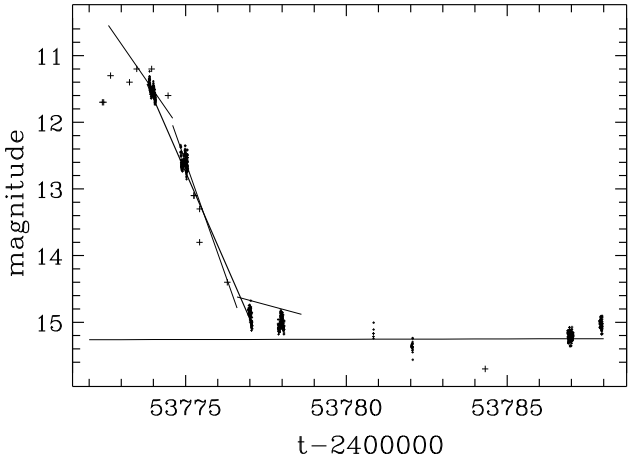}
\label{f5}
\caption{The composite light curve based on the AFOEV and VSOLJ international databases near the outburst. The inclined thin lines at the outburst show linear fits to the observations in the intervals $(-2^{\rm d},0^{\rm d}),$ $(0^{\rm d},2^{\rm d})$ and $(2^{\rm d},4^{\rm d}),$ respectively. The thick line is a linear fit for 3 nights of our R observations during a decay from an outburst. The bottom line corresponds to a part of a periodic $(P\approx303^{\rm d})$ sine fit to the "out of outburst" observations.}
\end{figure}

On a timescale of $\delta\tau=10^d,$ the light curve, shown in Fig. 4, is asymmetric, sharply rising before undergoing a gradual decay. This behavior is characteristic of a dwarf nova outburst (cf. Warner 1995, Hellier 2001). The total duration of the outburst does not exceed $5^d,$ and the ascending branch lasts $<1^d.$
A more reliable estimate could be calculated using the light curve (digitized graph) based on unpublished VSNET observations of the 1996 outburst (Kato, 1996): $T_{rise}=-{\rm d}t/{\rm d}m=0\fd085=122$min.

The mean brightness at the outburst maxima is $V=10\fm64\pm0\fm09$ (total range 9\fm7-11\fm2). Taking into account the mean quiescent value $V=15\fm17\pm0\fm01,$ the mean amplitude is 4\fm5. According to the ``Amplitude-cycle" statistical relationship (Kukarkin and Parenago 1934), the cycle corresponding to this amplitude, is $210^{\rm d}.$ Using improved coefficients of this relation for dwarf novae (Richter 1986), the cycle length was estimated to be $91^{\rm d}.$

The scatter in the data presented in Fig. 3 of Richter (1986), is considerable: the cycle length, corresponding to an amplitude $A=4\fm5$, as for other dwarf novae, is between $16^{\rm d}$ and $630^{\rm d}.$.

Another important characteristic of outbursts is a decay time $T_{decay}={\rm d}t/d{\rm m}$ (cf. Bailey 1975, Warner 1995), which is also called a ``decline rate" (cf. Smak 1984).
Kato et al. (2002) presented a statistical relation $\log (T_{decay}/1^d)=-0.25+0.79\log (P_{orb}/1^h),$ where $T_{decay}$ is characteristic time of decreasing brightness by $1^m.$ This is an improved relation, which was originally found by Bailey (1975). Assuming $P_{orb}=0\fd16537398$ (Haswell et al., 1997), $T_{decay}=1\fd67$ (or, looking at the scatter at the diagram at corresponding period, from $\sim1\fd1$ to $\sim2\fd6).$

The slope of the descending branch of the outburst light curve is changing. We have split the interval of $\tau_{ik}$ into bins. Visual inspection argues for binning with a width of $2^{\rm d},$ i.e. $(-2^{\rm d},0^{\rm d}),$ $(0^{\rm d},2^{\rm d}),$ $(2^{\rm d}, 4^{\rm d}).$ The values of $T_{decay}$ are 1.14(4) and 0.56(5) days/mag for the first two intervals, respectively. In other intervals, this parameter $T_{decay}$ seems to be not well defined. Approaching quiescence, the slope ${\rm d}m/{\rm d}t$ gradually decreases, so the light curve is very non-linear. Close to outburst maximum, the slope gradually increases. Various maxima
occur with different magnitudes and at different relative times $(\tau_{ik}).$ For the observed outbursts, the maxima occur at $\tau_{ik}\approx-2\fd1\pm0\fd4$ (ranging from $-3.6^{\rm d}$ to $-1^{\rm d}$).

Three nights of our CCD-R observations cover a descending branch of the outburst in 2006 (Fig.5). The corresponding value of $T_{decay}=0.902\pm0.003$ days/mag is an intermediate between values derived using visual observations. Generally, our R curve is in a good agreement with smoothed visual observations. This value obtained for the CCD R data for only one outburst, is in an excellent agreement with the value 0.92 days/mag (\v{S}imon 2000) obtained for a ``mean outburst" in V.
This measurement of the decay time is significantly smaller than expected from the improved Bailey's (1975) relation. Our result is consistent with the theoretical expectation of Angelini \& Verbunt (1989) for an accretion disk with an inner edge disrupted by the magnetic field of the white dwarf. As one can see from Figs. 2 and 3 of Angelini \& Verbunt (1989), a larger radius of the inner edge of the accretion disk leads to shorter burst and smaller decay rate.

Although there are two types of dwarf nova outbursts due to the disk instability starting at the {\em outer} (type A) or {\em inner} parts, as defined by Smak (1984), for both of them, the duration of the outburst decreases with an increasing inner radius of the disk (Angelini \& Verbunt 1989). So, from the single value of the decline time, one cannot distinguish between the two models (A or B), but the observational value argues that the presence of the magnetic field is required in both models.

The visual data (from the AFOEV and VSOLJ databases) shown in Fig.4, indicate that one night of our observations is located at the beginning of the descending branch. This outburst is one of the weakest (V=11\fm2) observed (the brightest outburst (V=9\fm7) occurs at JD 2448160).

Similar values $T_{decay}=0.84$ days/mag may be estimated from the light curve of the outburst in November, 2000, which was published by Szkody et al. (2002). A smaller value $T_{decay}=0.641(2)$ was determined for the outburst in 2005 from the observations of H.Maehara (AFOEV, 2007).
For the brightest historical outburst mentioned above, the ``2-day" slope is $T_{decay}=2.03\pm0.15$ after the maximum, much closer to the theoretical expectation. This may be due to the larger accretion rate and thus decreasing magnetosphere, so relatively ``non-magnetic" status of the system at the outburst maximum.

\subsection{Apparent Decrease of the Outburst Duration}

Unexpectedly, the value of $\tau_{ik}$ for the observed outbursts decreases with time $t.$ The correlation coefficient $\rho=0.97\pm0.11$ $(8.6\sigma)$ for ($t,$ $\tau_{ik}$) is very large. This implies that the observed outbursts become more shorter with time. Although such a correlation is apparently statistically significant, it could be created by a selection effect (being acquired more frequently, allowing the detection of more, short outbursts). This is true only partially, as, during the recent decade, the monitoring was regular. But the value of $(-\tau_{ik})$ (duration from the maximum to the level of $12^{\rm m},$ which is also some characteristic of the width of the outburst) has decreased from 2\fd9 to 1\fd0.

A physical explanation may be a precession of the magnetic white dwarf and corresponding variability of the Alfv\'en radius $R_A,$ which is dependent either on the accretion rate $\dot{M},$ or on the effective inclination of the magnetic axis to the orbital plane. This may cause a cyclic (or, observationally, non-monotonical) variability of the spin period of the white dwarf, which is observed in some intermediate polars, e.g. BG CMi (Kim et al. 2005b), FO Aqr (Andronov et al. 2005) and may be caused by precession of the white dwarf.

In this case, the observed correlation corresponds to a (temporarily) nearly linear part of (possibly, cyclic) variations of the outburst width. The decreasing width corresponds to an increasing inner radius of the disk (i.e. $R_A)$ (cf. Angelini \& Verbunt 1989).

In contrast to our new result on DO Dra,
the prototype of non-magnetic dwarf novae SS Cyg exhibits variations of the outburst cycle at half-century timescale. Another parameter, the width of outbursts, shows variations at shorter (a dozen years) and longer (more than a century) timescales (Fig. 14 of Cannizzo 1993).

Both a selection effect and a physical explanation may contribute to the detected apparent phenomenon. For two recent outbursts,  $(-\tau_{ik})\approx1^{\rm d}.$ Such short outburst durations make outbursts more complicated to detect. In practice, it needs a monitoring (at least twice per night) at one observatory and at least 3 observatories at various longitudes (e.g. East Asia, Europe, America). An excellent example of a ``non-stop" monitoring during 3 months was completed for WZ Sge after its superoutburst in 2001 (Ishioka et al. (2002), Patterson (2002)). DO Dra may be recommended for a similar monitoring (at least every 4 hours) to determine statistics of extremely narrow outbursts and study the underlying physics.

\subsection{"Decay Time - Outburst Maximum" Relationship}

\begin{figure}
\psfig{file=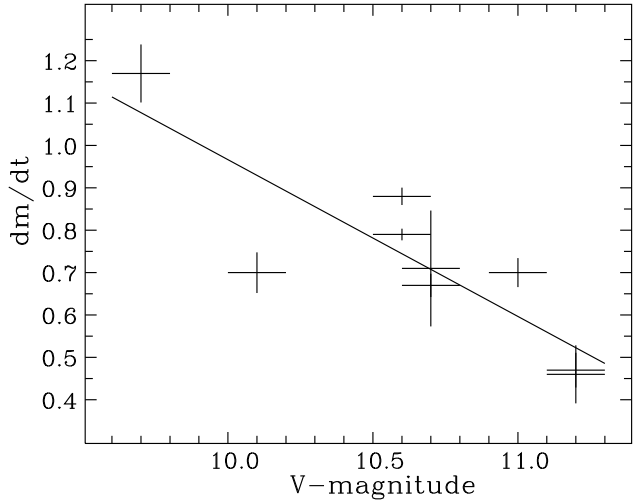}
\label{f6}
\caption{Dependence of the characteristic time of decline d$t$/d$m$ close to the adopted mean level of 12\fm5 on the brightness of the outburst based on the AFOEV and VSOLJ international databases.}
\end{figure}

Significant variability in the height and depth of the observed outbursts suggests changes of other parameters, e.g. the decay time $T_{decay}={\rm d}t/{\rm d}m$ estimated at the descending branch using nearby points  with a magnitude around a constant value of $12^{\rm m},$ mainly in the range $11^{\rm m}-13^{\rm m}$ (systematically brighter or fainter than the fixed value).
 The dependence of this parameter on magnitude V at maximum is presented in Fig. 6. One may see an excellent correlation (apart from one point) with a regression line:
${\rm d}t/{\rm d}m=0.73(4)-0.37(9)\cdot(V-10\fm64),$ i.e. the slope is equal to $4.3\sigma$ and thus is statistically-significant. One may note that the dependence of $-\tau_{ik}$ on V may be achieved, even if the form of the outburst remains the same, but it changes in height. Such a dependence appears to be obvious, so we do not analyze it further. However, the changes of slope correspond to changes in the outburst width, so this statistical dependence is interesting point to consider for future modeling.

According to theoretical models, assuming in a constant accretion rate $\dot{M},$ the inner $R_{in}$ and outer $R_{out}$ radii of the accretion disk, and the decay time $T_{decay}$ should be identical from outburst to outburst (Smak 1984, Angelini \& Verbunt 1989). For the models of outbursts in non-magnetic systems, $T_{decay}$ is constant, although the brightness at maximum varies drastically (cf. Fig. 3 in Smak 1984). The duration is smaller for smaller outbursts.

The observed difference of $T_{decay}$ for different outbursts argues for variability of parameters. The character of the dependence (decrease in $T_{decay}$ with decreasing brightness at maximum) is in agreement with model calculations. One would expect that a decrease in the accretion rate will cause an increase in the radius of the magnetosphere $R_A$ (e.g. Davidson \& Ostriker 1973, Lipunov 1992), then, consequently, a decrease of both $T_{decay}$ (Angelini \& Verbunt 1989) and the energy of the outburst.

At the upper part of the outburst, $T_{decay}$ is much larger than at the intermediate part of the descending branch. In a case of the constant shape of the outburst and a vertical shift at the light curve, this will cause an {\em increase} of $T_{decay}$ with decreasing maximum brightness instead of the observed {\em decrease}. This means that the observed statistical relation is dominated by outburst-to-outburst variability of $T_{decay}$ rather than by variability of the slope of the descending branch.

\section{Orbital-Scale Variability}

\begin{figure}
\psfig{file=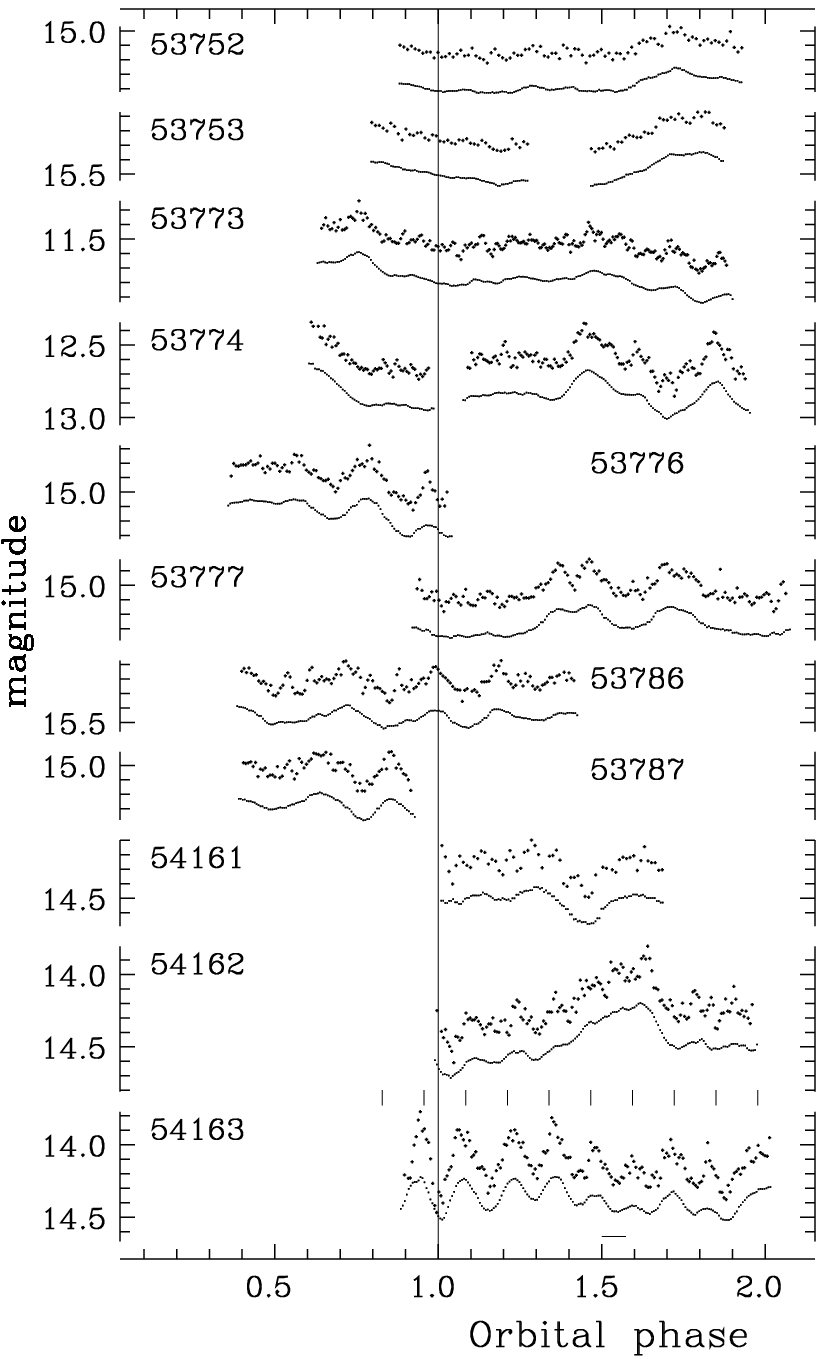}
\label{f7}
\caption{The individual light (R) curves for our observations. The numbers show integer parts of (HJD-2400000) for first observations at individual runs. The abscissa is expressed as an orbital phase according to Haswell et al (1997). The original data are plotted once and are not repeated once per unit phase. Below each original light curve (shifted by 0\fm25) is a "running sine" fit $a_0(t_0,\Delta t).$
For the last night with apparent high-amplitude brightness oscillations, the short vertical lines show positions of maxima according to the ephemeris for 30.4-min period. The short horizontal line below all curved shows the full width $(2\Delta t)$ for a "running sine" fit.
}
\end{figure}

\begin{figure*}
\psfig{file=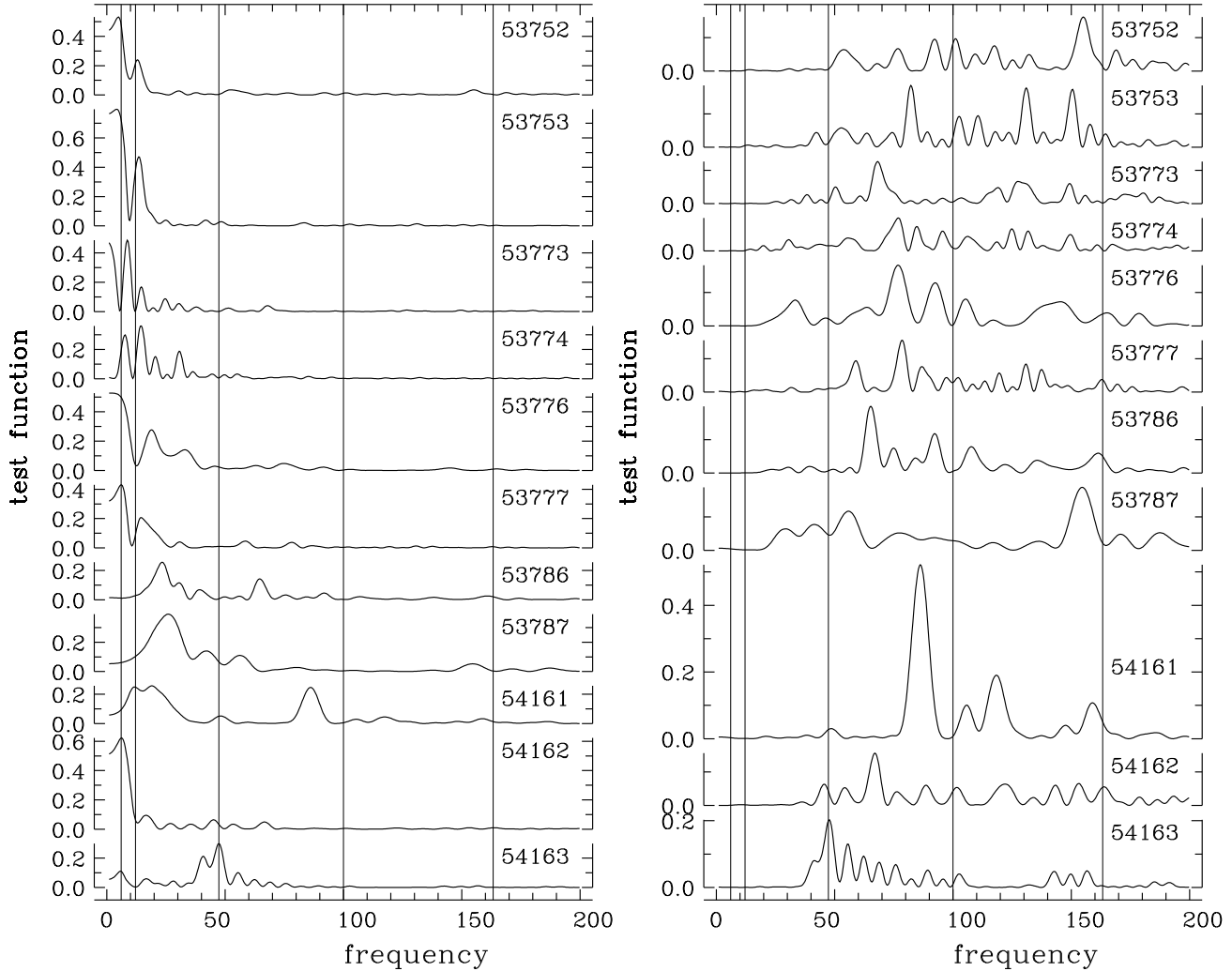}
\label{f8}
\caption{The periodograms $S(f)$ for original (left figure) data and for the residuals from the "running sine" trend (right figure). The numbers show integer parts of (HJD-2400000) for first observations at individual runs. The abscissa is expressed in cycles/day. The vertical lines show some suggested periods (from left to right): $P_{orb},$ $2P_{orb},$ the period 30.4 min best seen in the last night, the period 14.4 min seen during one night by Mateo et al. (1991) and the spin period (Szkody et al. 2002).
}
\end{figure*}

Although there are no eclipses in the system, there are variations with a peak-to-peak amplitude of
$0\fm17$ in H  and ``no sign of an orbital variation exceeding $\sim0\fm2$ in the ultraviolet" (Patterson et al. 1992). However, the light curves show variability of amplitude of $0\fm38-0\fm47$ for $\sim10-$minute averages, as one may estimate from their Fig.10. Mateo et al. (1991) pointed out, that the phase of the deepest minimum in B occurs close to the spectroscopic phase 0 (i.e. inferior conjunction of the donor star). In previous studies, an aperiodic flickering, in addition to a periodic component, was detected.
In Sect. 5, we present our analysis, using our own observations, to detect variability at higher frequencies.

Haswell et al. (1997) determined a most precise ephemeris
\begin{equation}
T_E=HJD 2446863.4376(5)+0.16537398(17)\cdot E,
\end{equation}
which we use to calculate orbital phases. Their mean phase curve is double-peak with an amplitude of 0\fm14 in I. The initial epoch 2449500.6585(2) for the deepest minimum also occurs close to phase 0. In other words, the phase 0 corresponds to minima in different spectral regions from ultraviolet to infrared, but the sources may be of different origin: the infrared curve may be mainly modulated due to distortion of the secondary, which fills its Roche lobe, whereas the ultraviolet emission is dominated by the accretion flow.

Drastic variability of the shape of the orbital curve was reported by other authors and is also seen in our data. The orbital curve is highly disturbed by other types of variability. Thus a periodic trigonometrical polynomial fit is not expected to produce good results. We followed instead the procedure of the ``running mean" with a rectangular filter of half-width $\Delta t=0\fd02$ proposed by Boyd (2004), who detected variability in V with an amplitude of $\sim0\fm2$.

The statistical properties of running fits with an arbitrary filter (weight) and basic functions, were studied by Andronov (1997). For the expected spin frequency $f=163$ cycles/day and the adopted rectangular  filter, the amplification coefficient is $(1-H(f\Delta t))=1-{\rm sinc\,}(1.63)=1.18$ (see Andronov 1997 for more details). In other words, the observed amplitude should be divided by this factor to obtain an unbiased estimate. Moreover, 18 per cent of the amplitude of the periodic signal will remain in the smoothing curve with an opposite sign. An alternative fit is a ``running sine" fit, which may be written in the following form:
\begin{equation}
m(t,t_0,\Delta t)=a-r\cdot\cos(2\pi((t-T_0)/P_0-\varphi)).
\end{equation}
Here $a(t_0,\Delta t)$ is the mean of the smoothing curve (which is generally different from the sample mean), $r(t_0,\Delta t)$ is the semi-amplitude and $\varphi(t_0,\Delta t)$ is the phase of the curve maximum, which corresponds to the initial epoch $T_0$ and initial period $P_0.$ For this approximation, additional ``rectangular" weights $p(z)$ are used, which are dependent on $z=(t-t_0)/P_0.$ The final smoothing function for $t=t_0$ is $m_C(t,\Delta t)=m(t,t,\Delta t).$

For the usual rectangular filter, $p(z)=1$ for $|z|\le1$ (i.e. $t_0-\Delta t\le t\le t_0+\Delta t),$ the Gauss function $p(z)=exp(-cz^2)$ corresponds to the Morlet wavelet (Foster 1996, Andronov 1998a), and Andronov (1997) proposed an intermediate function $p(z)=(1-z^2)^2$. The last function is continuous, contrary to the rectangular function, and local, contrary to the Gaussian. A comparison of the statistical properties of these fits was presented by Andronov (1998b). This method was applied e.g. to another intermediate polar BG CMi (Kim et al., 2005b).

For the analysis of our data, we have chosen a rectangular filter, as in Boyd (2004).
To decrease the distortion of the fit, it is highly recommended that integer values of the  the ratio ``filter width / period", are used.
Thus we have used $\Delta t=P_{spin}.$

We have compared the fits using the ``running mean", ``running line" and ``running sine" for the real data, and for simulated sine data with a period equal to $P_{spin}$ and unit amplitude at the moments of real observations. As expected, the running sine provides the best fit for the ``period-averaged" value of signal, whereas the running mean produces erroneous scatter about zero with a peak-to-peak amplitude of up to 20\% of the amplitude of the simulated sine signal. Taking into account better statistical properties, we therefore prefer to use the running-sine method, instead of the running-mean method applied by Boyd (2004).
 Boyd (2004) observed the star at the outburst stage, so the exposure was shorter, than in quiescence, and thus the scatter due to discreteness of data could be smaller, if applying $\Delta t=1P_{spin}$ instead of $\Delta t=1.63P_{spin}$ (latter value used by Boyd (2004)). Our observations span both maximum and minimum brightness and the exposure time differs for different brightness. The values of $m_k-a_0(t_k,\Delta t)$ will be called hereafter the ``residuals" (or ``detrended signal"), where the non-linear trend $a_0(t_k,\Delta t)$ was computed using the running parabola approximation (i.e. $a(t_0,\Delta t))$ with $\Delta t=1P_{spin}.$ As an initial value, $P_{spin}$ was set to the value $529\fs32=0\fd00612627$ (Szkody et al. 2002). The initial epoch was arbitrarily set to zero (i.e. HJD 2400000).

The individual light curves for 11 long nights are shown in Fig. 7. The remaining 3 runs are much shorter than the orbital period, and are therefore not shown. Fortunately, 3 subsequent nights monitored the descending branch of the light curve after outburst (Fig. 5).

The periodograms for 11 ``long" nights  of observations are shown in Fig. 8. The test function is $S(f)=1-\sigma^2_{O-C}(f)/\sigma^2_{O},$ where $\sigma^2_{O}$ and $\sigma^2_{O-C}(f)$ are variances of the original data, and of their residuals from a least squares one-harmonic approximation at a given trial frequency $f$ (see Andronov 1994 for details).


The highest peaks usually occur at low frequencies. Only for 4 nights from 11, this happens at the orbital frequency (53752, 53753, 53777 and 54162. However, no prominent regular brightness minima, which are close to the zero orbital phase, are observed at our light curves.

Using the ephemeris of Haswell et al. (1997), we reanalyze the published light curves in Fig. 10 of Patterson et al. (1992). Close to the ``computed" values of minima (corresponding to the minimum in I), at the U curves by Patterson et al. (1992), relatively wide minima are seen at HJD 2448268.95, 271.93, 4270.94. For these curves, the mean count rate was smaller by a factor of 1.3-1.5 than during the last night, when the ephemeris position HJD 2448272.922 corresponds to a ``narrow" minimum. Such type of variability is also seen in our data, when the minima are shifted by $(0.3-0.5)P_{orb}$ from the ephemeris.
Other explanations are that the additional emission arising during the outburst, has its maximum at phases opposite to the main maximum, or there may be an occasional distortion of the orbital light curve by a precessing accretion disk (like in SU UMa-type stars).

At the descending branch after the outburst (53774), there are 3 peaks with separations between them of $0.9P_{orb}$ and $0.4P_{orb},$ so there are 2 apparent peaks in the periodogram, although such photometric maxima seem not to be periodic.

Following discussions of the dependence of the amplitude on luminosity (cf. Szkody et al. 2002), we have estimated amplitudes of variations $\Delta R$ of the "non-spin" variability (i.e. $a_0(t_k,\Delta t)),$ additionally removing the trend. For the maximum of the outburst (JD 2453773, $\bar{R}=11\fm53),$ $\Delta R=0\fm23$ is distinctly smaller than for the following night (descending branch, $\bar{R}=12\fm60),$ when $\Delta R=0\fm33$ and is comparable with the amplitude $\Delta R=0\fm26$ at $\bar{R}=14\fm93.$ Here $\bar{R}$ is the mean brightness between minima and maxima. So the amplitude of the "non-spin" variability is largest at intermediate luminosity.

Other (last) group of 3 subsequent nights correspond to an ``intermediate" brightness, which slowly increased from $\bar{R}=14\fm27)$ to $14\fm13.$ This stage corresponds to the recent wide brightness maximum (Fig. 1). This is another state of activity, which is different from a rapid outburst, and has not been observed before.
Formally, $(T_{rise}\sim14)$ days/mag (determined from 3 nights only). During this state, the amplitude varied from $\Delta R=0\fm25$ to $0\fm27-0\fm43$ (depending on how the apparent trend was taken into account) and finally $0\fm30.$

In the ``quiescence" state $(15\fm02\le\bar{R}\le15\fm22),$ the amplitude ranges from $0\fm18$ to $0\fm24,$ i.e. is smaller than during both observed events of activity (i.e. the outburst and the brightening to the ``intermediate" level).

\section{Quasi- vs Transient Periodic Oscillations}

\begin{table}
\caption{Characteristics of the highest peaks at the periodogram $S(f)$ for the residuals $\tilde{m}_k:$ frequency $f$ (cycles/day), amplitude $r$ and their error estimates, period $P$(days), false alarm probability FAP.}
\begin{tabular}{cccccrrr}
\hline
JD&$f$&$\sigma_f$&$r$&$\sigma_r$&$P$&FAP\\
\hline
53752& 155.1& 0.8& 0.0150& 0.0037& 0.00645& $10^{-1.8}$\\
53753&  82.2& 0.7& 0.0132& 0.0032& 0.01216& $10^{-1.8}$\\
53773&  68.1& 0.5& 0.0155& 0.0027& 0.01468& $10^{-5.0}$\\
53774&  76.9& 0.5& 0.0182& 0.0039& 0.01300& $10^{-2.9}$\\
53776&  76.8& 1.1& 0.0238& 0.0055& 0.01302& $10^{-2.4}$\\
53777&  78.5& 0.5& 0.0228& 0.0044& 0.01274& $10^{-3.9}$\\
53786&  65.2& 0.6& 0.0254& 0.0044& 0.01533& $10^{-4.9}$\\
53787& 154.6& 1.7& 0.0227& 0.0058& 0.00647& $10^{-1.8}$\\
54161&  86.2& 0.6& 0.0599& 0.0076& 0.01160& $10^{-7.8}$\\
54162&  67.0& 0.7& 0.0319& 0.0061& 0.01492& $10^{-4.0}$\\
54163&  47.7& 0.4& 0.0425& 0.0064& 0.02096& $10^{-7.0}$\\
\hline
\end{tabular}
\end{table}

\begin{table}
\caption{Characteristics of the best cosine fits of the residuals assuming a fixed period $P_{orb}:$ initial epoch of maximum brightness $T_0$ and semi-amplitude $r$ their error estimates. The symbol ":" marks very bad approximation with fixed frequency.}
\begin{tabular}{cccccrrr}
\hline
$T_0$&$\sigma_T$&$r$&$\sigma_r$&Rem.\\
\hline
53673.0173 &0.0005& 0.0172& .0094&  \\
53753.0583 &0.0009& 0.0043& .0040& :\\
53754.0400 &0.0006& 0.0052& .0035&  \\
53774.0621 &0.0015& 0.0019& .0029& :\\
53775.0584 &0.0013& 0.0030& .0041& :\\
53777.0575 &0.0005& 0.0105& .0059&  \\
53778.0627 &0.0004& 0.0110& .0046&  \\
53780.8415 &0.0008& 0.0640& .0453& :\\
53782.0583 &0.0017& 0.0084& .0147& :\\
53787.0430 &0.0004& 0.0128& .0048&  \\
53787.9629 &0.0008& 0.0077& .0064& :\\
54161.9971 &0.0006& 0.0186& .0106&  \\
54163.0410 &0.0003& 0.0188& .0064&  \\
54164.0403 &0.0010& 0.0068& .0071& :\\
\hline
\end{tabular}
\end{table}

Quasi-Periodic Oscillations (QPOs) at characteristic timescales of dozens of minutes are often observed in cataclysmic variables, e.g. TT Ari (Tremko et al. 1996, Andronov et al. 1999). They are also present at our light curves. Because the periodograms for the original data (Fig. 8, left) are dominated by peaks, which correspond to low-frequency variability, we have recomputed the periodograms for the ``residuals" $m_k-a_0(t_k,\Delta t).$ They are shown in the right part of Fig. 8. The characteristics of the highest peaks in the periodograms for separate nights are listed in Table 4.

The most significant oscillations occurred during the last night of observations, when the mean magnitude was 14\fm13(1), a magnitude brighter than at quiescent level. From the periodogram analysis of the original data, we determined the period $P=0.02111(15)$ min, semi-amplitude $0\fm101(11)$ and an initial epoch $T_0={\rm HJD}\,2454163.9492(4).$ The positions of maxima according to this ephemeris are shown in Fig. 7. The amplitudes of individual brightness oscillations can differ by a factor of $\sim2.$ It should be noted that the amplitude estimate listed in Table 4 is formally much smaller $(0\fm043).$ This apparent result is due to a small frequency of oscillations, so the fit $a_0(t_0,\Delta t)$ follows the oscillations at a reduced amplitude. So, for this night, we prefer to use an ``unbiased" value $r=0\fm101(11).$ A similarly corrected amplitude for the previous night, is $r=0\fm047(11).$

In two previous nights, the oscillations are not seen so clearly in Fig.7, but they are also "periodic". The word  ``periodic" is used here to mean that, as for any other cataclysmic variable, all characteristics of the light-curve, cosine-function fit, its mean, amplitude and phase (and thus period), are variable, but the phase changes are much less than unity.
Contrary to ``periodic" variations, in many systems are observed ``quasi-periodic oscillations", where there the phase (and ``period") changes are drastic.

This episode of almost periodic oscillations during three consecutive nights at an intermediate brighness level is very interesting. Contrary to ``quasi-periodic oscillations" (QPO), we refer to this type of variability as ``transient periodic oscillations" (TPO). It is noticable that the amplitude changes during these 3 nights in a sequence of $0\fm06,$ $0\fm05,$ $0\fm10,$ i.e. not monotonically, whereas the frequency decreases monotonically  (86, 67, 47 cycles/day). The last period of 30.4(2) minutes apparently coincides with the double period of 14.4(2) min detected by Mateo et al. (1991).

One possible interpretation of these results is that we are observing a bright cloud rotating around a white dwarf. Assuming that the corotation radius is equivalent to the Alfv\'en radius $R_A$ (cf. Lipunov 1992), the 30-min period $(=3.45P_{spin})$ corresponds to the distance $R_{cloud}=R_A\cdot3.45^{2/3}=2.28R_A.$ This estimate is independent from the mass of the white dwarf. At such a distance, it is not expected that the cloud is disturbed by the rotating magnetosphere of the white dwarf. Adopting the mass ratio $M_2/M_1=0.45(5)$ (Haswell et al. 1997), we calculate a ratio of the distance to the orbital separation of $R_{cloud}/a=0.224(3),$ deeply within the Roche lobe.

Another interpretation is that a rotating cloud is being disturbed by the rotating magnetosphere of the rotating white dwarf (cf. the model of Hollander and van Paradijs (1992) for TT Ari). In this case, the oscillations should have a beat period $P=P_{cloud}P_{spin}/(P_{cloud}-P_{spin}),$ i.e. $P_{cloud}/P_{spin}=1/(1-P_{spin}/P)=1.41$ and $R_{cloud}/R_A=1.26,$ much closer to unity than in the first hypothesis. For the first night, the frequency is 86.2 cycles/day, so $R_{cloud}/R_A=1.65.$ The magnetic disturbance is expected to be effective at
$R_{cloud}/R_A\underline{\sim}1,$ so the calculated value of 1.65 argues against this interpretation.

The alternate hypothesis, contrary to that of the magnetic field distortion, is that the oscillations of brightness are due to modulated irradiation from a rotating cloud. In this case, the monotonic decrease of frequency agrees with a general picture of a cloud, which is spiralling towards a white dwarf. This suggestion should still be confirmed using by the Doppler imaging (see Schwope 2001 for further details). TPOs are however transient, and thus DO Dra may be proposed as a ``target of opportunity" for future studies at large telescopes.

\section{Spin variability}

The currently adopted value of the spin period is $P_{spin}=529\fs31(2)$ (Szkody et al. 2002).

Mateo et al (1991) determined a period of 14.2 minutes $(852^{\rm s})$ for one night of observations in B, which is significantly longer than other estimates. This value may also correspond to transient periodic oscillations. A single night of observations is, however, of limited use when trying to measure variations of the period.

It is interesting to note, that, contrary to other intermediate polars, there are no published reports on the O-C analysis of the spin variability. This task is difficult, because of the high spin frequency $f=163.231(6)$ cycles/day $\sim 59456(2)$ cycles/year (Szkody et al. 2002). Moreover, the wave form is unstable, making a precise phasing doubtful.

In Fig. 8, there are peaks at the spin frequency only during a few nights. During other nights, the nearby peaks are significantly shifted. Thus the phase light curves show minor amplitude, as one can see from Table 5. Athough the upper limit of the amplitude of $0\fm02$ is in agreement with previous estimates (Szkody et al., 2002, Boyd 2004), the shifts of the peaks from the spin frequency make estimates of the amplitude at the fixed frequency smaller than at the frequency corresponding to the peak.

Such a displacement is present sometimes in periodograms published in previous works, with a dominating frequency far from the "spin" one (cf. Patterson et al. 1991, Mateo et al. 1991). There is also no stable peak corresponding to the period 275s detected by Patterson et al. (1991) from the UV observations. Welsh and Martell (1996) reported on unusually blue spectral energy distribution of 529-s oscillations $f_{\nu}\propto(\nu/\nu_0)^{\alpha}$ with an $\alpha=2.2-2.8$ based on observations from optical to ultraviolet. This may be a cause of the small amplitude of 529-s oscillations measured in our R observations.

For comparison, the amplitude of the spin variations in in the filter R is only slightly smaller than in the filter V in other intermediate polars, e.g. BG CMi (Kim et al. 2005a), MU Cam (Kim et al. 2005b).
However, as was mentioned above, a prominent variability is seen in R at other frequencies.

\section{Results}

\begin{itemize}
\item From 14 nights of CCD R observations, 3 nights cover the descending branch of the outburst in 2006, from which a characteristic decay time $T_{decay}=0.902(3)$ days/mag was estimated.
\item The outburst cycle ranges from $311^{\rm d}$ to $422^{\rm d},$ contrary to a recently published estimate of $870^{\rm d}.$ However, due to the short duration of the outburst $\le 5^{\rm d},$ some outbursts are missing at the visual curves of the AFOEV and VSOLJ databases. An indirect evidence for this timescale is a photometric wave with a cycle $\sim 303(15)^{\rm d},$ which was detected in ``out of outburst" (quiescent data).
\item The profile of a ``composite outburst" is complicated, with $T_{decay}$ changing from 1.14(4) to 0.56(5) for $2^{\rm d}$ intervals before and after crossing of the constant brightness of $12^{\rm m}$. The previous time interval, which contains an ascending branch, corresponds to a wide variety of outburst height and duration.
\item A strong correlation was found between the decay time and the maximum brightness of historical outbursts. With decreasing brightness, the decay time also decreases, in contrast to the expectations of the model of ``magnitude shift" of the constant shape. This may be interpreted by variability of the inner disk edge radius, which is the radius of magnetosphere, and is consistent with the disk-instability calculations of Angelini \& Verbunt (1989) (see their fig. 2 and 3).
 Additionally, the extrapolation of this statistical relationship may argue for the presence of a number of missed weak narrow outbursts.
\item In common with other authors, our observations do not show any periodic variability with an orbital phase. The amplitude of the smoothing function ranges from $0\fm18-0\fm24$ in the ``quiescence" state, and increases to $0\fm33$ on the descending branch after the outburst. At the outburst peak, the amplitude was only $0\fm24.$ The amplitude therefore appears to be maximal in the intermediate brightness state.
\item During 3 subsequent nights in 2007, we have detected a type of variability, which was not previously reported, which we propose to name ``transient periodic oscillations" (TPO). During each night, a well-defined periodicity was detected, even though the frequency changed from night-to-night, and decreased monotonically from 86 to 47 cycles/day, and the semi-amplitude increased from $0\fm06$ to $0\fm10.$ At the same time, the system was by $\sim1^{\fm m}$ brighter than at the quiescence, and its brightness monotonically increased from $14\fm27$ to $14\fm13.$ This decrease in frequency is consistent with the model of a beat character of this variability, particularly, caused by irradiation of some cloud, which is ,for example, spiralling down to the white dwarf, its frequency increasing and approaching the spin frequency. It should be noted, that in 2007, the mean brightness had shown a long shallow peak lasting till autumn,
during which no (expected) outburst was detected. The TPOs were observed only during this prolonged state of activity. To study this new interesting phenomenon, new photometric and spectral observations are required.
\end{itemize}

\begin{acknowledgements}
This research was supported by the Korea Astronomy Observatory and Space Science Insititue Research Fund 2007 and was partially supported by the Ministry of Education and Science of Ukraine.
 In section 2, this research has made use of the AFOEV databases operated at CDS (France), VSOLJ and VSNET databases (Japan) and NASA's Astrophysics Data System Abstract Service. We thank Dr. Elena Pavlenko (Ukraine), Dr. Taichi Kato (Japan) and Pavol Dubovsky (Slovakia) for fruitful discussions and an anonymous referee for helpful comments.
\end{acknowledgements}

\end{document}